# Hydrodynamic 3D Simulation of Roche Lobe Overflow in High-mass X-Ray Binaries

David Dickson
Department of Physics, North Carolina State University Raleigh, NC 27695-8202, USA

## Abstract

While binary merger events have been an active area of study in both simulations and observational work, the formation channels by which a high-mass star extends from Roche lobe overflow (RLO) in a decaying orbit of a black-hole (BH) companion to a binary black-hole (BBH) system merits further investigation. Variable length-scales must be employed to accurately represent the dynamical fluid transfer and morphological development of the primary star as it conforms to a diminishing Roche lobe under the runaway influence of the proximal BH. We have simulated and evolved binary mass flow under these conditions to better identify the key transitional processes from RLO to BBHs. We demonstrate a new methodology to model RLO systems to unprecedented resolution simultaneously across the envelope, donor wind, tidal stream, and accretion disk regimes without reliance upon previously universal symmetry, mass flux, and angular momentum flux assumptions. We have applied this method to the semidetached high-mass X-ray binary M33 X-7 in order to provide a direct comparison to recent observations of an RLO candidate system at two overflow states of overfilling factors $f = 1.01$ and $f = 1.1$. We found extreme overflow ($f = 1.1$) to be entirely conservative in both mass and angular momentum transport, forming a conical L1 tidal stream of density and deflected angle comparable to existing predictions. This case lies within the unstable mass transfer (MT) regime as recently proposed of M33 X-7. The $f = 1.01$ case differed in stream geometry, accretion disk size, and efficiency, demonstrating nonconservative stable MT through a ballistic uniform-width stream. The nonconservative and stable nature of the $f = 1.01$ case MT also suggests that existing assumptions of semidetached binaries undergoing RLO may mischaracterize their role and distribution as progenitors of BBHs and common envelopes.

*Unified Astronomy Thesaurus concepts:* X-ray transient sources (1852); Stellar winds (1636); Stellar accretion disks (1579); High mass x-ray binary stars (733); Stellar mass black holes (1611); Hydrodynamical simulations (767); Astrophysical fluid dynamics (101); Roche lobe overflow (2155); High energy astrophysics (739)

## 1. Introduction

Roche lobe overflow (RLO), the process by which a donor star overfills its Roche lobe and loses mass to an accretor companion, is vital to the evolution of close binary systems and can determine the resultant novas, mergers, and/or remnants those systems produce (S. H. Lubow & F. H. Shu 1975; K. Pavlovskii et al. 2017; P. Marchant et al. 2021; A. S. Bunzel et al. 2023). G. Duchene & A. Kraus (2013) found at least 80% of O-type stars have at least one stellar companion. H. Sana & C. J. Evans (2010) found most of the surveyed O-type stars in binaries were in short-period orbits ($P < 100$ days). Therefore a significant portion of O-type stars may be close enough that RLO or wind RLO (in which only the stellar atmosphere overflows the Roche lobe) could influence their lifespans, novae, and remnants (N. Ivanova et al. 2013). RLO in semidetached binaries with a compact-object accretor forms the most promising progenitor of binary black holes (BBHs), which are critical to gravitational-wave astronomy (J. Frank et al. 2002; N. Ivanova et al. 2013; B. P. Abbott et al. 2016; S. S. Bavera et al. 2021; A. S. Bunzel et al. 2023; A. Dorozsmai & S. Toonen 2024). RLO also serves as the sole progenitor of common envelope (CE) systems and the exotic remnants they can produce (N. Ivanova et al. 2013). Due to their high rate of mass transfer (MT), the timescale of RLO-induced MT onto black hole (BH) accretors is expected to be very brief, limiting observational data of this vital progenitor process (S. H. Lubow & F. H. Shu 1975; P. Marchant et al. 2021; V. Ramachandran et al. 2022).

The accreting BH binary M33 X-7 has provided a rare observational window into the RLO phase in high-mass X-ray binaries (HMXBs; V. Ramachandran et al. 2022). The spectral analysis of V. Ramachandran et al. (2022) concludes M33 X-7's donor star to be significantly overfilling its Roche lobe, supplying the accretor with the excess mass required to generate its high X-ray luminosity (Table 1). This differs from the previous analysis of J. A. Orosz et al. (2007), which reported that the donor star fueled the BH's X-ray luminosity with wind accretion alone as it did not fill its Roche lobe. They derived a filling factor $f \equiv R_{\mathrm{donor}}/R_{\mathrm{RL,vol}} \approx 0.77$, where $R_{\mathrm{RL,vol}}$ is the volume-equivalent radius of the donor's Roche lobe (J. A. Orosz et al. 2007). V. Ramachandran et al. (2022) predict M33 X-7's MT phase is in the process of becoming unstable and subsequently may form a CE, rather than a BBH as its currently stable MT phase would suggest (K. Pavlovskii et al. 2017). By simulating the system dynamics of M33 X-7, we intend to capitalize on this rare glimpse of RLO onto a BH accretor to improve parameter-space constraints on this elusive process. Detailed examination of M33 X-7 may reveal broader implications about three parameters as of yet unconstrained by existing research: MT rate, MT efficiency, and tidally fed angular momentum loss.

Despite its evolutionary relevance, the RLO MT rate has not yet been constrained as a function of donor radius or the associated overfilling factor $f$. The rate, and therefore timescale, of RLO-induced MT is not only an observational limitation but an unknown parameter of its own (J. Frank et al. 2002). The MT rate is expected to vary by its stability, which is critical to the subsequent multiphase evolution of close binary systems (K. Pavlovskii et al. 2017; P. Marchant et al. 2021; A. Dorozsmai & S. Toonen 2024). Recent work has begun to relate both MT rate and stability to donor radius through spherically symmetric 1D radial simulations and analytical methods (P. Marchant et al. 2021). Stable MT, characterized by an expanding or constant donor Roche lobe, describes the regime in which net negative feedback occurs in MT. This causes the donor star to be stripped of only its outer envelope, leaving the donor near its Roche lobe throughout a long evolutionary period and potentially forming a BH. Unstable MT describes the opposite case, in which net positive feedback strips ever more of the expanding star away from its diminishing Roche lobe (J. Frank et al. 2002; K. Pavlovskii et al. 2017; P. Marchant et al. 2021). P. Marchant et al. (2021) determined donor stars with radiative envelopes that enter stable MT to be the primary progenitors of BBHs, while other donors were more likely to generate CEs. They arrived at these conclusions assuming HMXB RLO phase is a 1D hydrostatic system with a steady, transonic, adiabatic, and purely radial tidal stream subject to no stellar rotation or tidal coupling (P. Marchant et al. 2021). Hydrostatic simulations such as this neglect key elements of the system dynamics that could influence the stability of MT and impact results obtained (S. H. Lubow & F. H. Shu 1975; P. Marchant et al. 2021).

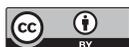

**Table 1**
M33 X-7 System Parameters

| Parameter | Symbol | Value |
| --- | --- | --- |
| Mass of Accretor Black Hole | $M_{\mathrm{accretor}}$ | 11.4 $M_\odot$ |
| Mass of Donor Star | $M_{\mathrm{donor}}$ | 38 $M_\odot$ |
| Orbital Period | $P_{\mathrm{orb}}$ | 3.45 days |
| Binary Separation | $D$ | 35 $R_\odot$ |
| Temperature of Donor Star | $T_{\mathrm{eff}}$ | 31 kK |
| Terminal Velocity of Donor Wind | $V_\infty$ | 1500 km s$^{-1}$ |
| Mass Loss Rate of Donor Wind | $\dot{M}_{\mathrm{wind}}$ | $4.0 \times 10^{19}$ g s$^{-1}$ |
| Stellar Radius of Donor Star | $R_{\mathrm{donor}}$ | 20.5 $R_\odot$ |
| Hydrogen Mass Fraction of Donor Star | $X_{\mathrm{H}}$ | 0.60 |
| Helium Mass Fraction of Donor Star | $X_{\mathrm{He}}$ | 0.395 |

The efficiency of this binary system MT flux, $\alpha_{\mathrm{MT(tot)}} = \dot{M}_{\mathrm{accretor}}/-\dot{M}_{\mathrm{donor}}$, has yet to be critically examined in either the stable or unstable MT regime by hydrodynamic simulations. Similarly absent from existing study is the relationship between the MT efficiency of the binary system and that of its L1 tidal stream, which may predominate accretion. We calculate stream MT efficiency by comparing the L1 mass flux to the dense outflow stream generated by its nonconservative entry onto the accretion disk, $\alpha_{\mathrm{MT(L1)}} = 1 - \dot{M}_{\mathrm{stream,out}}/\dot{M}_{\mathrm{L1}}$. These efficiencies play a pivotal role in subsequent system evolution. In the case of BBH formation, greater system mass from conservative MT shortens inspiral and heightens gravitational-wave amplitude. A. Olejak et al. (2021) recently found modern techniques used to synthesize BBH populations to be vulnerable to insufficiently constrained RLO physics. That same vulnerability, born of simplistic assumptions made of the efficiency and rate of RLO-induced MT, limits the efficacy of all models that rely on RLO evolution (S. H. Lubow & F. H. Shu 1975; P. Marchant et al. 2021; A. S. Bunzel et al. 2023).

Angular momentum, as well as mass, is lost from the system by inefficient MT. However, simple models and prescribed MT simulations often neglect this possibility. CE formation and initial BBH separation are dictated by angular momentum loss during the RLO phase (N. Ivanova et al. 2013). In the seminal work of S. H. Lubow & F. H. Shu (1975), the angular momentum carried by MT has been assumed equivalent to that of a stable orbit of the center of mass that passes through the L1 point. They also assume fully conservative MT and neglect angular momentum transfer within the accretion disk, eliminating the possibility of MT-induced angular momentum loss (S. H. Lubow & F. H. Shu 1975). With these assumptions, they concluded the MT tidal stream to launch at an angle of 19°.5–28°.4 into a uniform-width column for much of its length before tapering on its curved approach to the accretion disk (S. H. Lubow & F. H. Shu 1975). They also provide prescriptions of stream width, density, and accretion disk radius. Despite the underlying assumptions that neglect angular momentum loss, these conclusions have been widely used as prescriptions of L1 MT.

Simulations relying on prescribed L1 MT, rather than a fully realized donor star, have become standard in the study of RLO-fed accretion (R. Whitehurst 1988; Z. Meglicki et al. 1993; S. Kunze et al. 2001; A. Zhilkin et al. 2019, 2022). Donor envelope dynamics have primarily been modeled separately using the hydrostatic stellar evolution code MESA (B. Paxton et al. 2010; F. Valsecchi et al. 2015; K. Pavlovskii & N. Ivanova 2015; M. Renzo et al. 2023; S. Gossage et al. 2023). The spherical symmetry of MESA precludes the asymmetric deformation of an overflowing envelope; MESA donor simulations therefore rely on prescriptive MT rates or indirect inferences rather than self-consistent hydrodynamical tidal stream simulation (B. Paxton et al. 2010). To constrain RLO mass and angular momentum flux, and their associated transfer efficiencies, multidimensional hydrodynamics are needed in modeling the complete binary system.

When multidimensional hydrodynamic simulations of complete RLO binaries have been attempted through grid solutions or smoothed particle hydrodynamics (SPH), they have been constrained to low resolution. SPH tracks individual particles through system contours, enabling varied length and timescales to be examined on a particular mass scale with $10^4$–$10^6$ particle resolutions for 3D RLO (V. Nazarenko & L. Glazunova 2003; C.-P. Lajoie & A. Sills 2010; N. de Vries et al. 2014; T. A. Reichardt et al. 2019). Grid solutions must reconcile varying length scales, limiting the effectiveness of uniformly distributed grid zoning to constrain small-scale flow. For 3D RLO, even well-designed nonuniform grids have been almost exclusively low resolution, spanning $10^5$ zones (D. Bisikalo et al. 1997, 1998a, 1998b, 1998c, 1999, 2000; K. Oka et al. 2002). V. Nazarenko & L. Glazunova (2006) achieved a landmark nonuniform Cartesian grid resolution of $10^6$ zones. Rather than resolving the accretion process, they still assume 50% of incident mass entering the accretion disk ultimately accretes onto the compact object.

## 2. Methods

We employ 3D hydrodynamic simulations to constrain the mass and angular momentum loss from the donor, and associated efficiencies of accretion onto the compact companion, in HMXBs undergoing RLO. Our simulations are built with the VH-1 code, which uses the piecewise parabolic method of P. Colella & P. R. Woodward (1984) to solve Euler's equations for an ideal gas with an adiabatic index $\gamma = \frac{5}{3}$ (J. Blondin & A. Taylor 2024, in preparation). VH-1 conserves angular momentum and enacts radiative cooling in 3D nonuniform grids, making it ideal for our purposes (J. Blondin & A. Taylor 2024, in preparation). We have modeled M33 X-7 using a set of spherical grids in VH-1, following the method laid out by J. Blondin & A. Taylor (2024, in preparation) for the Vela X-1 system. We utilize one set of spherical grids centered on the donor and likewise on the accretor, computed in the frame corotating with the binary system. This grid-based code enables us to simulate the donor envelope, photosphere, and wind simultaneously with a high-resolution model of the accretion flow onto the compact companion.

We have brought this conceptual framework into a well-optimized nonuniform set of paired grids of unprecedented resolution. We employ four grids of 512 radial zones, 96 polar increments, and 288 azimuthal increments, combined into two overset spherical yin-yang configurations to achieve an angular resolution of 1° across the entire simulation window (A. Kageyama & T. Sato 2004; J. Blondin & A. Taylor 2024, in preparation). The accretor is mapped by a set of high-resolution grids housed inside the larger donor grids and paired across boundary zones. Within the donor envelope and photosphere, the donor grid resolution is set to maintain a steady well-resolved flow. We define this resolution by half of the photospheric scale height $SH/2 \equiv c_s^2 R_{\rm donor}^2/(2GM_{\rm donor}) =$ 7114 km. The donor grids extend to larger zones in the wind regime where scale lengths are significantly longer. The accretor grids are similarly nonuniform, with innermost zones of $\Delta R = 70.6$ km and zones scaling larger with radius, maintaining a constant $\Delta R/R$ for approximately cubic zones with respect to the angular resolution, $\Delta R = R\Delta\theta$. To fully span the accretor Roche lobe, we chose an accretor-centered grid outer limit $\approx 10\,R_\odot$. We model the accretor grids down to the inner range of the accretion disk, limited by our code's slow angular momentum transfer and lack of X-ray feedback from the accretor. The accretor grids therefore span radially from $0.12\,R_\odot$ to $10.6\,R_\odot$. The donor grids span radially from $13.3\,R_\odot$ to $52.8\,R_\odot$ to encapsulate the accretor grids accounting for the binary separation. Our grid dimensions of $512 \times 96 \times 288 \times 2 \times 2$ span a total of $5.7 \times 10^7$ zones. With these improvements to resolution, our complete binary grid enables us to examine the MT rate, its associated efficiency ($\alpha_{\rm MT}$), and angular momentum loss, each insufficiently constrained by previous simulations.

In order to model RLO in cases like M33 X-7 in which the donor radius significantly exceeds the Roche surface, we have added an outer envelope to the numerical method of J. Blondin & A. Taylor (2024, in preparation). The initial outer envelope is constructed as a spherically symmetric radiative envelope and then mapped as a function of gravitational potential to the effective potential of the binary system. This solution is computed assuming a constant Kramer's opacity and the stellar parameters mass, radius, and luminosity as given in Table 1 for M33 X-7 (L. F. Shampine 1975; R. Kippenhahn et al. 1990). In keeping with the method of J. Blondin & A. Taylor (2024, in preparation), we treat both donor and accretor as point sources for gravitational calculations, though our envelope contains a distributed mass of $\sim 0.01 M_\odot$ on grid in the 1.1 case. The 1.01 case envelope maps only $\sim 0.0002 M_\odot$ on grid.

For continuity, we then pair the envelope to the wind profile modeled to the limits of our grid resolution (Figure 1) at $R_{\rm donor} = 20.5\,R_\odot$ following observations of V. Ramachandran et al. (2022). This is consistent with a photospheric density (optical depth $\tau = 1$) of $5.72 \times 10^{-10}$ g cm$^{-3}$ in keeping with the model O II atmosphere of surface gravity $\log g = 3.4$ and $T_{\rm eff} = 31$ kK presented by R. Hainich et al. (2019). We observed wind launching to supersonic velocities at densities of $1.4 \times 10^{-12}$ g cm$^{-3}$ from the exponential-density isothermal atmosphere. In the 3D model, the envelope is initially identical on any radial slice to the contours given in Figure 1, except mapped by aspherical effective potential rather than radius.

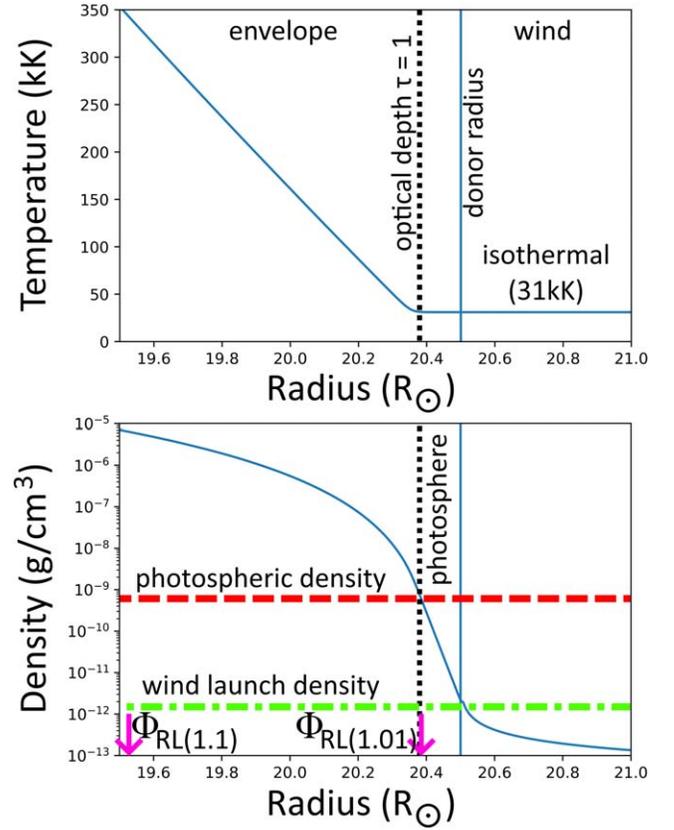

**Figure 1.** 1D envelope-to-wind transition plotted in temperature and density. Photospheric density corresponds to an optical depth $\tau = 1$, in keeping with the model O II atmosphere of R. Hainich et al. (2019). Wind launch density indicates where velocities become supersonic. Pink arrows labeled by $\Phi_{\rm RL(1.01)}$ and $\Phi_{\rm RL(1.1)}$ demarcate the effective potential offset of the Roche lobe relative to the donor radius for both 1.01 and 1.1 cases.

To map our solution onto the effective gravitational potential contours of an aspherical overflowing star with $f > 1$, we must define their characteristic radii. While volume-equivalent radii are generally favored for radii at or below the Roche potential, this breaks down for non-CE overflowing donor stars. We instead chose to define the donor surface and Roche lobe by their eclipse radii, $R_{\rm donor} = R_{\rm donor,ecl}$. We determine eclipse radius with the equipotential surface truncated by the plane passing through the L1 point orthogonal to the line connecting the centers of masses of the system. Without this choice of truncation, any RLO equipotential surface of $R_{\rm donor} > R_{\rm RL,ecl}$ fully encloses the accretor, though this equipotential is only completely filled in CEs. We chose to simulate the system at two thresholds of overflow. In the more extreme overflow case, we chose the envelope surface to match an eclipse radius of $R_{\rm donor} = 20.5\,R_\odot$ in agreement with V. Ramachandran et al. (2022). This yields an overfilling factor $f \equiv R_{\rm donor}/R_{\rm RL,ecl} = 1.10$ and thus we call it the 1.1 case. The 1.1 case most closely corresponds to the observations of V. Ramachandran et al. (2022) for direct comparison, but a second, less overflowing case is needed to compare with broad RLO models, such as the work of S. H. Lubow & F. H. Shu (1975) and P. Marchant et al. (2021). We therefore examine a second model at an eclipse radius $R_{\mathrm{donor}} = 18.7\,R_\odot$. We call this slightly overflowing model the 1.01 case, with an overfilling factor $f = 1.01$. We then map densities and pressures by corresponding offset potential to fill the donor star region ($R \leqslant R_{\mathrm{L1}}$). This solution is then paired to the wind as we did in the 1D case. We chose to only initialize the envelope within $R \leqslant R_{\mathrm{L1}}$ to avoid inverted density gradients beyond L1. This necessitates a brief settling period (time evolution) for the L1 flow to reach a steady state.

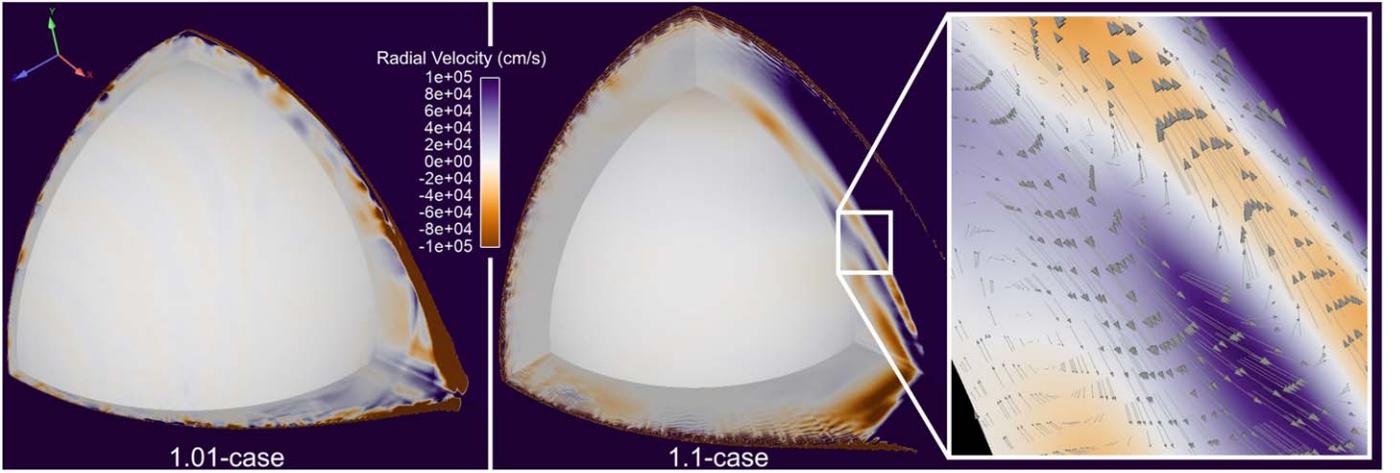

**Figure 2.** Stellar interior mapped by radial velocities. For both cases, the inner boundary of the grid is mapped by the central region and the ecliptic is set in the upper right quadrant. In the ecliptic plane, the deep tidal-feeding flow structure forms in contrast to the relatively quiet poles. Deep eddy flows that emerge from the 1.1 case are further detailed by velocity vectors in the ecliptic plane. Velocities are all significantly subsonic, with sound speeds $c_s \geqslant 6 \times 10^6$ cm s$^{-1}$ everywhere in the stellar interior.

We model the donor envelope to a depth sufficient to be characterized by quasistatic, highly subsonic motion with negligible radial flux; this ensures the envelope mass flows that feed the tidal stream are steady and physical. This inner boundary gradually supplies mass to the grid to maintain constant pressure in the innermost grid zones. Due to the negligible radial flux near the boundary, this flow does not substantially vary the system dynamics other than to refill the donor envelope as mass overflows and is lost through L1; system dynamics were insensitive to a complete cutoff of the inner boundary flow on timescales briefer than days, though simulations of greater duration saw gradual reduction in overflow as the envelope diminished.

We therefore simulate the full extent of relevant mass during the MT phase and enable mass upwelling into the outer regime stripped by MT. Our model initializes with spherically symmetric winds corresponding to a Sobolev-driven $\beta$-law velocity profile and both donor photosphere and envelope corotating with the system (D. G. Hummer & G. B. Rybicki 1985; R.-P. Kudritzki & J. Puls 2000). The tidal stream and disk are both absent from the initialization, instead allowed to form freely during the settling phase. The time evolution required for model settling occurs computationally quickly in all components other than the accretion disk. Figure 2 shows the radial velocity component within the envelope and particularly at the inner boundary. Perturbed surface flows developed in the photosphere and were unable to propagate deeper into the envelope due to the pressure and density gradient. The 1.1 case also exhibited deep eddies with complete turnover near the ecliptic, mapped by velocity vectors on the right-hand side of Figure 2.

Holistically modeling RLO requires careful mediation between disparate thermodynamical extremes. Deep within the stellar envelope, radiative transport should dominate the entropy profile with respect to effective potential (R. Kippenhahn et al. 1990). In contrast, the surface flows entering the overflowing tidal stream are sufficiently dynamic to transport energy adiabatically, and recent works have taken the stream as adiabatic as well (P. Marchant et al. 2021). The emergence of tidal-feeding flows upwelling from deep within the envelope in our simulation suggest the invalidity of that assumption; these deep flows lead to a steady rise in the entropy of the tidal stream across simulation time as gas is drawn up from ever deeper in the envelope. As we are unable to achieve radiative transport in a 3D grid solution, we instead maintain the deep envelope in a radiative steady state and implement a smooth numerical transition between the radiative profile and adiabatic dynamical flow.

Our numerical method begins by calculating the radiative and dynamical timescales to cross each individual zone within the envelope at each computational step. This calculation is shown in Equation (1), using $\Delta r$ to denote to the radial dimension of a particular grid zone and $\kappa$ to denote opacity.

$$t_{\mathrm{dyn}} = \Delta r / |\vec{v}|, \quad t_{\mathrm{rad}} = \kappa \Delta r^2 \rho / c \qquad (1)$$

We then apply a weighted average within each zone, weighting the pressure induced by unconstrained adiabatic flow against that of the initial radiative thermal gradient (Equation (2)). $P_{\mathrm{radiative}}$ corresponds to the pressure required to maintain equal temperature across the appropriate equipotential surface, while $P_{\mathrm{adiabatic}}$ is calculated by adiabatic dynamical evolution at each step. This allows a smooth transition from the radiative region to the dynamical flows induced by the tidal stream. Our weighted average is taken at each iterative step, which must be mediated by a factor of $dt/10^6$ s, selected based on the settling timescale of the tidal stream, so as not to bias dynamical evolution by varying

Table 2
Simulated Mass and Angular Momentum Fluxes

| Region | Property | Symbol | 1.01 case | 1.1 case |
|---|---|---|---|---|
| Donor Wind | Mass Loss Rate | $\dot{M}_{\rm wind}$ | $5.36 \times 10^{19}$ | $6.03 \times 10^{19}$ |
| | Angular Momentum Loss Rate | $\dot{L}_{\rm wind}$ | $1.67 \times 10^{39}$ | $2.03 \times 10^{39}$ |
| Donor L1 | Mass Loss Rate | $\dot{M}_{\rm L1}$ | $1.7 \times 10^{18}$ | $8.80 \times 10^{22}$ |
| | Angular Momentum Loss Rate | $\dot{L}_{\rm L1}$ | $3.1 \times 10^{37}$ | $1.18 \times 10^{42}$ |
| External | Mass Loss Rate | $\dot{M}_{\rm ext}$ | $5.14 \times 10^{19}$ | $5.84 \times 10^{19}$ |
| | Angular Momentum Loss Rate | $\dot{L}_{\rm ext}$ | $1.64 \times 10^{39}$ | $1.96 \times 10^{39}$ |
| L1 Tidal Stream | Mass Transfer Efficiency | $\alpha_{\rm MT(L1)}$ | 0.86 | 1.00 |
| | Angular Momentum Transfer Efficiency | $\alpha_{\rm AM(L1)}$ | 0.20 | 1.00 |
| Binary System | Mass Transfer Efficiency | $\alpha_{\rm MT(tot)}$ | 0.07 | 1.00 |
| | Angular Momentum Transfer Efficiency | $\alpha_{\rm AM(tot)}$ | 0.04 | 1.00 |

**Note.** Values expressed in cgs units where appropriate.

frequency of iterative steps.

$$P = \left( \frac{P_{\rm adiabatic}}{t_{\rm dyn}} + \frac{P_{\rm radiative}}{t_{\rm rad}} \frac{dt}{10^6 \text{ s}} \right) \times \left( t_{\rm dyn}^{-1} + t_{\rm rad}^{-1} \frac{dt}{10^6 \text{ s}} \right)^{-1}. \quad (2)$$

Radiative energy loss dominates over both of these considerations in the regime of the thin disk and optically thin wind. We define the wind regime as the region with densities below the photospheric density of the donor star, which inherently defines it as optically thin. We set a floor temperature on the near-accretor region such that the disk cannot thin to less than one grid zone in scale height; the high-density regime within this near-accretor region we define as the disk regime. The outer reaches of disk and stellar wind are therefore best characterized by an assumption of highly efficient radiative cooling that results in an isotherm (J. M. Blondin et al. 1990). We set them to a temperature $T_{\rm eff} = 31$ kK in keeping with observations (V. Ramachandran et al. 2022).

Our cool, thin disk model may not account for the full range of relevant energy transport dynamics. At sufficiently high overflow, this assumption of efficient cooling may break down; optically thick high-density disks may warm and generate greater outflows than we have modeled here. Recent observations also suggest M33 X-7 may be near its Eddington luminosity. If this is the case, the feedback ionization and radiation pressure could substantially differ from this prescription. The outer envelope may also have significant Helium abundance that could impact the system's response to X-ray feedback from BH accretion. X-ray heating and astrochemistry are outside the scope of this paper, but we hope to investigate these effects in future work.

### 3. Results

Our simulation methodology, applied to the parameters observed of M33 X-7 described in V. Ramachandran et al. (2022), has achieved a quasisteady hydrodynamic equilibrium within donor wind, donor envelope, and tidal flow in both 1.01 and 1.1 cases. Angular momentum transfer in RLO binaries, previously limited to analytical assumptions, is simulated here for the first time (Table 2).

Our 3D profile demonstrates a steady-state donor wind which averages $v_\infty = 1500$ km s$^{-1}$ and $\dot{M}_{\rm wind} \approx 5$ to $6 \times 10^{19}$ g s$^{-1}$ across all sightlines not perturbed by the tidal flow, in keeping with V. Ramachandran et al. (2022; Table 2). To resolve the launched donor wind, we examine its flux at the L1 radius. We exempt the tidal stream to be summed separately as the L1 mass flux (Table 2). The Sobolev-driven wind launches from the aspherical surface of the donor envelope, rapidly accelerating to higher velocities than those occurring within the stellar interior (Figure 2). The donor wind launches with angular momentum and mass fluxes almost identical to the quantities observed exiting our simulation's external boundary. These were constant across simulation time in both cases (Figure 3). This corresponds to a system angular momentum loss timescale of $\tau_{\dot{L}} \equiv L/\dot{L} \approx 20$ Myr in both cases.

The donor envelope solutions we present in Figure 2 are subsonic and radiatively dominated throughout, with the near-exception of the substantial 1.1-case ecliptic flow. The 1.01-case donor envelope developed significantly subsonic flows that die off quickly toward the interior. The deep envelope flows induced were sufficiently insignificant that we elected to alter the donor boundary conditions; by raising the inner boundary of the envelope, we were able to increase the resolution of the donor wind regime near the external boundary. We found the L1 and wind mass fluxes to vary by less than 1% with an increase in the donor-grid inner radius to 15.4 $R_\odot$. This elevated inner boundary is shown in Figure 2. This change in resolution did not result in a significant change in wind flux, though the lesser envelope depth of the 1.01-case model increased its computational efficiency.

The 1.1-case donor developed more dynamic flows in the ecliptic. A significant deep outflow emerged, which feeds the higher tidal stream density we observed accompanying the greater overflow. The dynamical nature of this deep outflow also generated eddy currents exhibiting complete turnover deep within the radiative envelope (Figure 2). The 1.1-case donor envelope was simulated to a greater depth sufficient to model the entirety of the tidal-feeding outflow. This flow exceeds Mach 0.6 near the surface deposition of the tidal-feeding stream and has dynamical timescales as brief as $\tau_{\rm dyn} \approx 10 \tau_{\rm rad}$, but remains radiative. In both cases, the stellar interior motion is entirely subsonic, unlike the tidal stream.

The L1 tidal stream differed significantly between the 1.01 and 1.1 cases, in both structure and scale. In both cases, we saw the tidal stream accelerate through L1 to the outer band of the accretion disk. That mass slowly accreted through our disk to the BH grid boundary (Figure 4). The accretion of the L1 tidal stream onto the outer edge of the disk forms an impact shock front that leads to a dense outflow stream beyond the disk. The material unbound from the system through the dense outflow stream is markedly similar in mass flux across both cases. However, the mass flux of the L1 stream itself differed greatly from $\dot{M}_{L1(1.01)} = 1.7 \times 10^{18}$ g s$^{-1}$ in the barely overflowing 1.01 case to $\dot{M}_{L1(1.1)} = 8.80 \times 10^{22}$ g s$^{-1}$ in the extreme-overflow 1.1 case (Table 2). The L1 tidal stream in the 1.01 case exceeds the donor wind in density but not in total mass flux, unlike the 1.1 case, which substantially exceeds the associated wind mass loss rate (Figure 3). This MT is conservative, with an efficiency $\alpha_{MT(1.1)} = 1.00$. The barely overflowing 1.01 case differed from the conservative model, with MT efficiency $\alpha_{MT(1.01)} = 0.86$ on average across the simulated time window. This discrepancy is a result of the relative invariance of external flows; much like the donor wind, the dense outflow stream remains of the same order in mass and angular momentum flux in spite of the ~50,000 times increase in $\dot{M}_{L1}$ between the two cases. In the 1.01 case, this outflow therefore represents a much more substantial proportion of the overflowing material than it does in the 1.1 case.

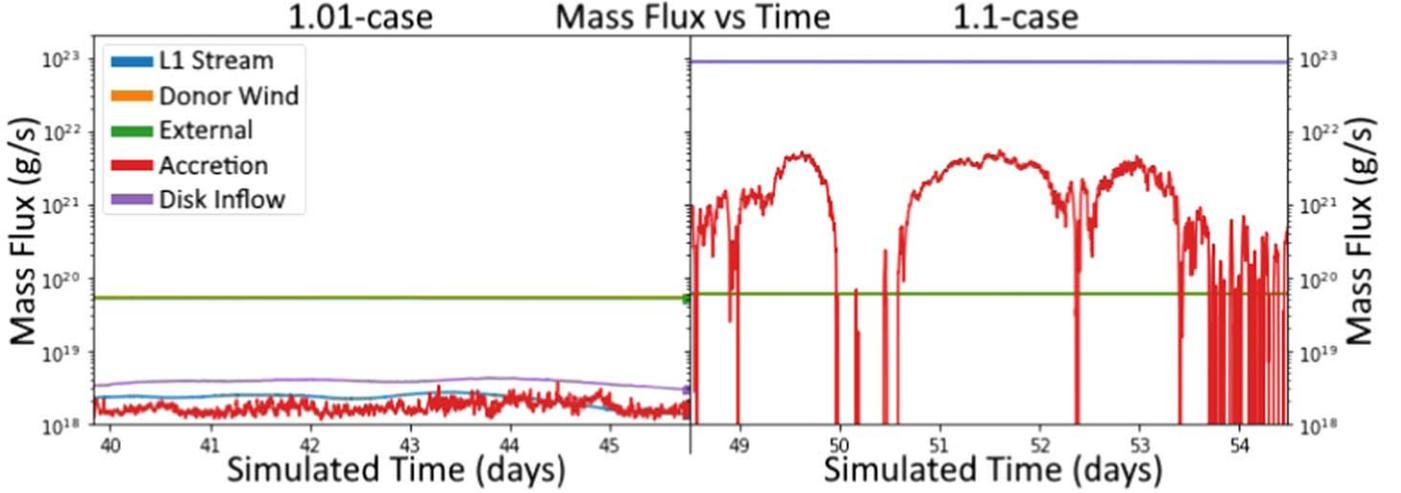

**Figure 3.** Mass flux through system zones. The L1 stream sums zones exceeding the maximal density of the supersonic wind at the L1 radius, $3 \times 10^{-13}$ g cm$^{-3}$. The lower mass zones on the spherical surface at the L1 radius around the donor are summed for the donor wind mass flux. Disk inflow corresponds to the net inflow through an accretor-centered spherical surface fully enclosing the disk. Accretion demarcates the net inflow through a spherical surface in the interdisk zone. The L1 stream and disk inflow mass fluxes of the 1.1 case are nearly identical across simulated time. The donor wind and external mass fluxes are nearly identical across simulation time in both cases. We exclude the settling time required to achieve a quasisteady state from initial conditions, in which fluxes were not generally observed nor physically meaningful.

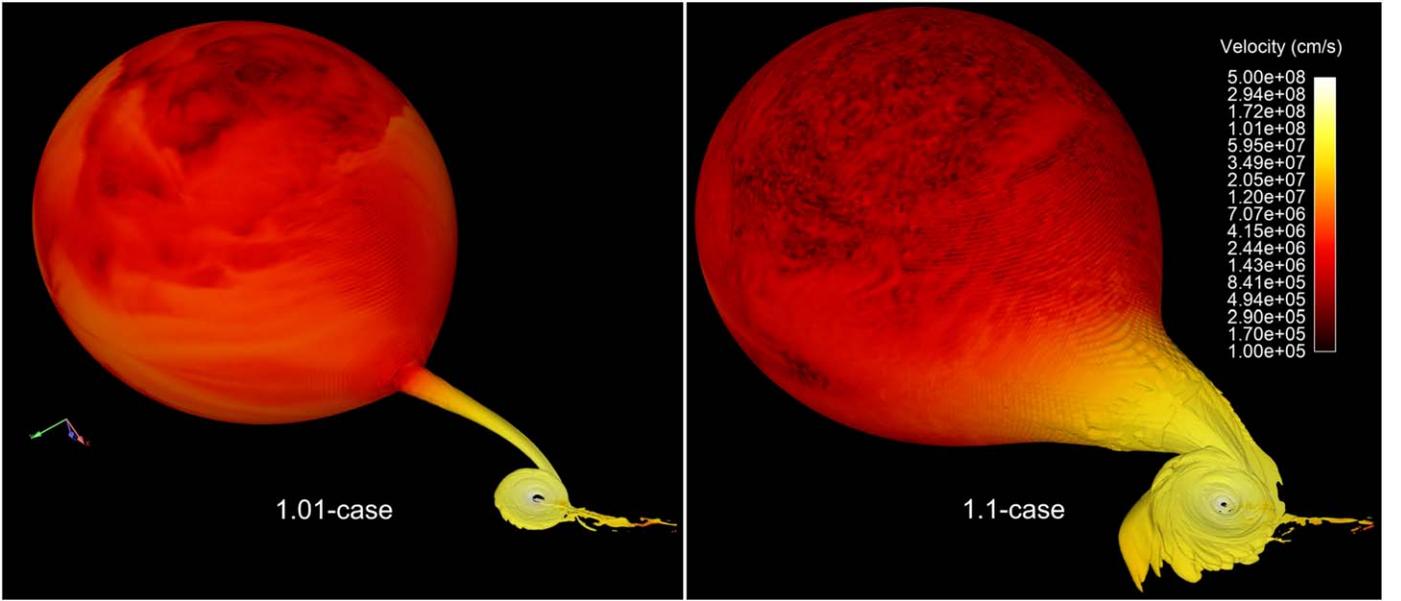

**Figure 4.** Donor, tidal stream, and banded accretion disk visualized by an isodensity surface. The tidal stream is angled with respect to the line connecting centers of mass, and is centered on the ecliptic. View given from below the ecliptic to better image the stream impact shock front and the nonconservative outflow stream formed where the tidal stream collides with the disk.

The invariance of the dense outflow stream may result from its reliance on the donor wind. The stream-disk interaction generates an outflow with initially insufficient energy to escape. Some of this outflowing mass is then driven by the donor wind deflected by the accretor, producing the outflow we observe. The driving effect of the constant wind appears unable to carry significantly more mass away from the 1.1-case system than the 1.01-case system, leaving more mass to fall back into the disk.

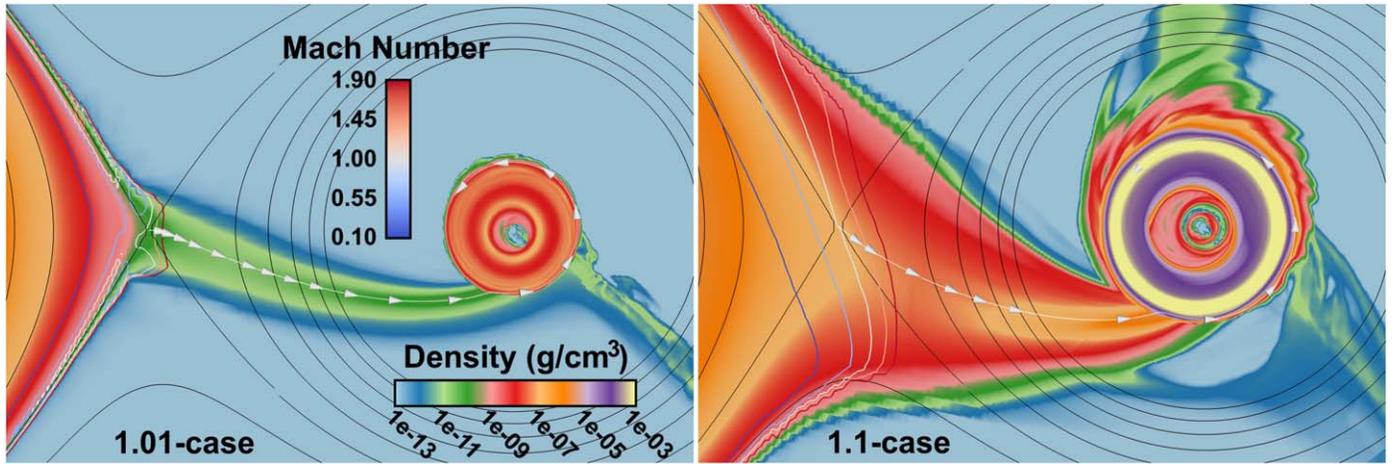

**Figure 5.** System ecliptic mapped with equipotential lines in black. Streamline arrows indicate flow pattern extending from the L1 point. Colored contours have been provided to map the transonic region. Accretion disk bands separated by the lower density gap are shown by density gradient.

Table 3
Comparison to Quantitative Predictions of Previous Work

| Property | Case | Our Data | Lubow and Shu | Marchant et al. | Ramachandran et al. |
|---|---|---|---|---|---|
| Tidal Stream Angle | Both | $17.5°$ | $20°$ | ... | ... |
| Tidal Stream Width | 1.01 | $4.2 R_\odot$ | $1.3 R_\odot$ | $1.4 R_\odot$ | ... |
|  | 1.1 | $12 R_\odot$ | $1.3 R_\odot$ | $10 R_\odot$ | ... |
| Donor L1 Angular Momentum Loss Rate | 1.01 | $3.1 \times 10^{37}$ | $3.2 \times 10^{37}$ | ... | ... |
|  | 1.1 | $1.18 \times 10^{42}$ | $1.67 \times 10^{42}$ | ... | ... |
| L1 Lagrange Point Density | 1.01 | $4.0 \times 10^{-10}$ | $1.12 \times 10^{-10}$ | $2.29 \times 10^{-10}$ | ... |
|  | 1.1 | $1.1 \times 10^{-7}$ | $5.94 \times 10^{-6}$ | $1.60 \times 10^{-7}$ | ... |
| Donor Wind Mass Loss Rate | 1.01 | $5.36 \times 10^{19}$ | ... | ... | $4.0 \times 10^{19}$ |
|  | 1.1 | $6.03 \times 10^{19}$ | ... | ... | $4.0 \times 10^{19}$ |
| Donor Wind Mass Capture Rate | 1.01 | $2.3 \times 10^{18}$ | ... | $3.49 \times 10^{18}$ | ... |
|  | 1.1 | $2 \times 10^{18}$ | ... | $4.38 \times 10^{18}$ | ... |
| Donor L1 Mass Capture Rate | 1.01 | $1.5 \times 10^{18}$ | ... | $2.27 \times 10^{18}$ | ... |
|  | 1.1 | $8.80 \times 10^{22}$ | ... | $3.61 \times 10^{22}$ | ... |
| Disk Outer Radius | 1.01 | $2.47 R_\odot$ | $4.9 R_\odot$ | ... | ... |
|  | 1.1 | $3.83 R_\odot$ | $4.9 R_\odot$ | ... | ... |
| Accretion Rate | 1.01 | $1.71 \times 10^{18}$ | ... | ... | $2.8 \times 10^{17}$ |
|  | 1.1 | $2 \times 10^{21}$ | ... | ... | $2.8 \times 10^{17}$ |

**Note.** Values expressed in cgs units except where otherwise specified. Mass capture rates indicate net mass entering the disk regime. Repeated values indicate predictions independent of overfilling factors or their associated L1 mass fluxes.

This would lead to the dense outflow stream increasing in total energy as it is driven out toward the grid boundary, as we observed.

In both the 1.01 and 1.1 cases, we observed the separation of the accretion disk into distinct concentric bands. In Figure 5, a visible separation forms between the tidal stream influx and the inner band of the accretion disk in both cases. Across a spherical shell separating the two bands, the mass flux contained within $\pm 5°$ of the ecliptic outweighs the off-ecliptic mass flux across both models. Despite the gap, this indicates that mass accretion is primarily driven by the tidal stream, not the donor wind. While our code can develop this banded disk structure, the computation time required to simulate the inner deposition onto the black hole is prohibitive. Additionally, observation of M33 X-7 suggests the innermost regime would be dominated by radiative feedback that our model does not account for (V. Ramachandran et al. 2022). Therefore, we simulate only to an inner bound of $0.12~R_\odot$ and assume all mass that crosses the disk gap ($R_{gap} \approx R_\odot$) is eventually accreted, which is to be taken as an upper limit of accretion onto an X-ray source BH.

## 4. Conclusions

While HMXBs and other contact binaries have been the subject of extensive theoretical and computational work, we here present the highest resolution model to date of both RLO and the associated tidally fed disk dynamics. Our simulation holistically models the donor envelope, photosphere, wind, and L1 tidal stream, as well as the accretion disk surrounding the BH accretor, on a nonuniform grid. We applied this model to the eclipsing HMXB M33 X-7 in two cases. We name them for their overfilling factors $f \equiv R_{donor,ecl}/R_{RL,ecl}$ as the barely overflowing 1.01 case and the greatly overflowing 1.1 case.

The tidal-stream geometry set out by S. H. Lubow & F. H. Shu (1975) is largely validated by our 1.01 case model, though the extreme overflow of the 1.1 case differs. In keeping with their semianalytical model, both our 1.01-case and 1.1-case tidal streams achieve Mach 1 precisely at the L1 point (Figure 5). We also found the angular momentum carried by the tidal stream to be on the order of the circular orbit through the L1 point about the system's center of mass, as predicted by S. H. Lubow & F. H. Shu (1975; Table 3). They also prescribe the angle of the tidal stream from the most direct line between the two bodies as between 19°.5 and 28°.4 across all mass ratios. For the mass ratio of M33 X-7, the predictions of S. H. Lubow & F. H. Shu (1975) differ from our observed angle by 2°.5. Shown in Figure 5, both models follow equivalent angles of deflection despite disparate overfilling factors, as predicted. Unlike the predictions of S. H. Lubow & F. H. Shu (1975), the 1.1-case tidal stream forms a conical funnel, rather than remaining of nearly uniform width across the nearly straight segment preceding its inspiral. This discrepancy is likely the product of their assumption that overflow occurs only in a small region about the L1 point; in the 1.1 case, the tidal-stream width at L1 was 12 $R_\odot$, of the same order as the binary separation. S. H. Lubow & F. H. Shu (1975) predicted stream width to be invariant with respect to L1 mass flux, which arises from their small-overflow-region assumption. For the parameters of M33 X-7, they prescribe the tidal stream to be of width 1.3 $R_\odot$. Both exceeded this value, to varying degrees which suggest a dependence, which suggests a dependence upon overfilling factor.

In both cases, a cross-section at L1 revealed the streams to exhibit circular symmetry with heights identical to their widths to the limits of our grid resolution. We observe donor envelope flows to be primarily constrained to the ecliptic except in the solid angle beneath the tidal stream in both cases. Significant outflows occur outside the ecliptic in this region, which is much larger in the high overflow case. These flows generate circular symmetry in the stream well away from the axis connecting centers of masses, where stream densities are substantially lower than at its center. These outflows remain shallower and slower than those observed in the ecliptic. This suggests the majority but not entirety of mass entering the tidal stream is first driven into the equatorial region prior to overflowing, as predicted by S. H. Lubow & F. H. Shu (1975).

The discrepancy of tidal-stream width correlates closely with a discrepancy in L1 density. While the 1.01 case had comparable density to their predicted value at the L1 point, our 1.1 case was significantly broader and less dense than their model would predict for the same mass flux. They also predict the outer radius of the accretion disk to be independent of tidal-stream mass flux, prescribing a disk radius of 4.9 $R_\odot$. This arose from the radius at which they predicted a ballistic stream would collide with itself after one orbit of the accretor. Both our stream widths and deflection angles differed from this geometry, which may account for the difference between their predicted value and the more compact disks we observed. As we do not examine angular momentum transport all the way into the BH, there exists a possibility that this transport would broaden the disk radii. However, this broadening is limited by the stream-disk interaction, which primarily expels angular momentum through the dense outflow stream, which carries a specific angular momentum of $1.1 \times 10^{20}$ cm$^2$ s.

Recent work by P. Marchant et al. (2021) provides an updated semianalytical method which generalizes beyond the small-overflow approximations of S. H. Lubow & F. H. Shu (1975) and presumes adiabatic tidal flow rather than isothermal; their method more closely fit our results of tidal-stream mass flux and geometry. Their method was accurate to within 1 order of magnitude in predicting the wind and L1 stream mass capture rates for both the 1.01 and 1.1 cases, where the mass capture rate is defined by the donor mass flux captured by the accretor (Table 3). We similarly validated their method with respect to the density at the L1 Lagrange point. Their methodology relies on the assumption of the L1 point being solely fed by surface flows along the Roche equipotential surface, which differed from our observations; this did not substantially impact the density obtained, though it may play a more substantive role at greater overflow than we studied here. Their method predicts tidal stream widths from the effective potential difference in the L1 plane. This method underestimates our findings but provides a substantially better fit than the S. H. Lubow & F. H. Shu (1975) prescription (P. Marchant et al. 2021). Our final point of comparison to the work of P. Marchant et al. (2021) is in their calculation and examination of MT stability.

MT stability is often reduced to a simple threshold of mass ratio, but we here substantiate that nonconservative dynamics can generate stable MT well beyond the customary mass ratio limit (J. Frank et al. 2002). MT stability in semicontact binaries is defined by the proportional rate of change in the volume-equivalent Roche lobe radius $R_{\rm RL,vol}$. If we assume fully conservative MT with no angular momentum loss modeled in isolation of the donor wind, as is customary, we may instead use a critical value of mass ratio (J. Frank et al. 2002). M33 X-7 has a mass ratio $q = M_{\rm donor}/M_{\rm accretor} = 3.3 \gg 0.83$, which would place this system well within the customary unstable MT regime (J. Frank et al. 2002; P. Marchant et al. 2021; V. Ramachandran et al. 2022). However, instead of using the $q \approx 0.83$ stability limit, we may more accurately calculate $\dot{R}_{\rm RL,vol}$ from the generalized analytic formula given in P. P. Eggleton (1983). This yields a Roche lobe shrinking timescale of 24 Myr in the 1.01 case, greatly differing from the 1.1 case timescale of 4200 yr. The 1.1-case timescale may be sufficiently short for corotating synchronicity to break down; if that is the case, use of the Roche geometry may not fully capture the system dynamics.

In the 1.01 case, the change in Roche lobe radius is driven primarily by the wind overflow, and its Roche lobe shrinks on a timescale of the same order as the total angular momentum loss and total mass loss of M33 X-7. This case therefore corresponds to stable MT occurring well outside of the customary mass ratio limit $q \approx 0.83$ (J. Frank et al. 2002). The 1.1 case's rapid Roche lobe evolution, driven by the high-overflow L1 tidal stream, places it in the unstable MT regime. The discrepancy of stability regime between the 1.01 and 1.1 cases validates the conclusion of V. Ramachandran et al. (2022) that the system is transitioning from stable to unstable MT. The stability of the 1.01 case, in contrast to the 1.1 case of identical mass ratio, suggests the mass ratio's invalidity as a predictor of system stability and progenitor pathway. The formation of a secondary MT stream at L2 that could significantly modify the stability regime has been suggested by P. Marchant et al. (2021); our simulations were unable to launch such a stream except under extreme overflow significantly beyond the 1.1 case.

Beyond stability, another side effect of RLO MT in our simulation was the enhancement of the wind. Though the wind remained of the same order, we saw a significant discrepancy in $\dot{M}_{\rm wind}$ between the 1D case and our two 3D cases, 1.01 and 1.1. We fit our 1D model to $\dot{M}_{\rm wind(1D)} = 4.0 \times 10^{19}$ g s$^{-1}$ as observed by V. Ramachandran et al. (2022). The same simulation parameters defined our two 3D cases, however they differed unavoidably. Due to the rotation of the primary, we saw slower, denser wind near the ecliptic and faster, less dense wind driven from the poles. We therefore discuss wind mass flux as a sum across the spherical surface that excludes the solid angle represented by the optically thick L1 tidal stream. With this method, our 1.01 case saw a ≈35% increase over the 1D wind mass flux. The 1.1 case saw an even greater ≈52% increase over the 1D case (Table 3). This is made all the more notable for the 1.1 case's wide L1 tidal stream excluding an even greater solid angle from wind contribution. Though recent work by J. Blondin & A. Taylor (2024, in preparation) has examined an analogous effect in the wind RLO regime, further research is required to separate the impact of 3D modeling wind asymmetries, the result of greater donor surface area in RLO systems, and the direct consequence of the L1 tidal stream on wind enhancement.

Our simulation provides novel constraints on the efficiency of mass and angular momentum delivery in RLO-fed MT. In keeping with the predictions of P. Marchant et al. (2021), we found the 1.1 case to have significant loss of angular momentum from the donor, which would result in orbital hardening without requiring a CE phase (Table 3). The results of both cases validated the prediction of S. H. Lubow & F. H. Shu (1975) that the L1 tidal stream carries angular momentum equivalent to that of a circular orbit of the center of mass that passes through the L1 point. While our simulation cannot resolve all the way to the BH accretor surface, we provide upper limits of mass accretion taken from the mass flux inflowing through the disk gap without X-ray feedback. Even at this upper limit, we found the 1.01 case did not exceed the Eddington luminosity. Our 1.01-case accretion averaged $1.7 \times 10^{18}$ g s$^{-1}$, or ≈73% of the Eddington value; this exceeds but is of the same order as the ≈12% finding of V. Ramachandran et al. (2022). Our 1.1-case accretion was orders of magnitude higher, ≈1000 times the Eddington value, suggesting an extreme rate of X-ray feedback would be required to reach equilibrium.

Future work may improve upon our method by applying additional physics to the dynamics and simulating a wider range of RLO systems. Our radiative envelope solution does not account for stellar evolution due to the short timescale of the simulation; future research could incorporate MESA or a similar stellar evolution code into system initialization to model varied stellar evolutionary stages (B. Paxton et al. 2010). At the transition from the isothermal photosphere to the linearly increasing temperatures of the envelope, we see winds launch naturally from the star. These winds follow the Sobolev approximation for linearly driven stellar winds, but we acknowledge the asymmetry of our donor star limits the applicability of this approximation. Further research could expand upon this by more physically constraining the wind dynamics. The greatest limitation of our work is its lack of accounting for X-ray feedback and relativity in the near-BH region. The applicability of this model could be expanded by incorporating a more complete radiative transport implementation. Particularly to near-Eddington and super-Eddington systems, the role of relativistic jets, X-ray heating, optically thick warm disk structures, and radiation pressure could significantly impact the accretion dynamics and stream-disk interaction. At the short timescale of the 1.1-case Roche evolution, the donor may not maintain corotation with the binary system. This could impact the effective potentials and therefore the MT rates, efficiencies, and geometry. Further research may improve constraints upon the high-overflow regime by examining the system without the assumption of a corotating donor.

## Acknowledgments

We are grateful for the guidance and mentorship of John Blondin. We acknowledge support from NSF grant AST-2308141.

## ORCID iDs

David Dickson  https://orcid.org/0009-0006-4860-9212